\documentstyle[12pt,epsfig]{article} 

\begin{document}

\begin{flushright}
\end{flushright}

\begin{center}
{\huge Momentum Resolution Improvement Technique for Silicon 
Tracking Detectors using dE/dx \\}
\vspace*{1.5cm}
{\large 
Stathes D. Paganis
\footnote{
Corresponding author present address: DESY-F1, ZEUS Group 
Notkestrasse 85, HH 22603, Germany.
e-mail address: paganis@nevis1.columbia.edu}\\
\vspace*{0.2cm}
{\small Columbia University, Nevis Laboratorties,\\
P.O.Box 137, Irvington NY 10533, USA}\\
\vspace*{0.5cm}
Jaw-Luen Tang
\footnote{e-mail address:  jawluen@bnl.gov}\\
\vspace*{0.2cm}
{\small Department of Physics,\\
National Central University, Chung-Li, Taiwan 320}

}

\end{center}
\vspace*{2cm}

\begin{abstract}
A technique for improving the momentum resolution for low momentum 
charged particles in few layer silicon based trackers 
is presented. The particle
momenta are determined from the measured Landau $dE/dx$ distribution 
and the Bethe-Bloch formula in the $1/\beta^2$ region.
It is shown
that a factor of two improvement of the momentum determination is 
achieved as compared to standard track fitting methods. This
improvement is important in large scale heavy ion experiments which 
cover the low transverse momentum spectra 
using stand-alone silicon tracking 
devices with a few planes like the ones used 
in STAR at RHIC and ALICE at LHC. 

\end{abstract}

\vspace*{2cm}

\noindent
PACS 29.40.Gx,
29.30.-h,
12.38.Mh

\newpage

\section{Introduction}
\vspace{0.2in}

Relativistic heavy ion experiments are performed in order 
to observe and prove the existence of 
a new form of matter, the Quark Gluon Plasma (QGP). 
A number of QGP observables rely on accurate measurement 
of the low transverse to the beam momentum 
($p_{\perp}$) particle spectrum. Such measurement 
can provide an accurate interaction source size 
(as obtained from pion correlations), 
better temperature information, improved 
strange and multistrange particle reconstruction.
For these reasons, large scale heavy ion experiments use vertex 
detectors and in particular silicon trackers, in order to deal with 
the high multiplicity of the collisions. STAR experiment at RHIC 
uses a 3 layer Silicon Vertex Tracker (SVT) and ALICE at 
LHC a 6 layer Inner Tracking System (ITS).
For particle momenta below $200 MeV/c$ there is significant increase  
of the multiple Coulomb scattering effect 
which results in a very poor 
momentum determination when standard track fitting methods are applied.
Despite the very good position resolution of the silicon trackers
(10-20 $\mu m$),
the momentum resolution is $dp/p \simeq 25\%$ for $p=100MeV$.

The purpose of this paper is to present an alternative method which 
can provide an improvement in the momentum determination by at least 
a factor of 2. The momentum is directly determined from the $dE/dx$ 
measurement of every reconstructed track using the mean measured
$\langle dE/dx \rangle$.
The reasons for this dramatic
improvement are: first, in the $1/\beta{^2}$ region for silicon the dE/dx
distribution is narrow and can be fitted by
a Gaussian with a small standard deviation, and second 
the slope of $\langle dE/dx \rangle (p)$ 
curve is large so that two neighboring
distributions $\langle dE/dx \rangle (p_1),
\langle dE/dx \rangle (p_2)$ have a small overlap.
Consequently one can use the mean $\langle dE/dx \rangle(p)$ 
of a track and the Bethe-Bloch curve to obtain a 
momentum with a relatively small error (around $10\%$).

The organization of the paper is as follows: in the first section we 
present an analytical proof of the method's validity, in the second 
section a simulation for a particular tracker is performed 
as an application of the method and finally the last section contains 
our conclusions.

\section{Improving the momentum determination}
\vspace{0.2in}

In this section we show analytically using the well known 
Bethe-Bloch formula and the Landau $dE/dx$ distribution for pions 
traversing a thin silicon layer, that the expected error 
in the the momentum determination of the incident pions 
is of order $10\%$ for $p < 200~MeV$. 

\noindent
We start with the Bethe-Bloch formula \cite{bethe}:
\begin{equation}
\frac{dE}{dx} = \frac{4\pi Ne^4}{mc^2\beta^2} z^2 
(ln\frac{2mc^2\beta^2\gamma^2}{I} - \beta^2) 
\label{eqn : bethe}
\end{equation}
where $mc^2$ is the rest energy of the electron, $z$ the charge 
of the travelling particle, N the number density of electrons in the 
matter traversed and $I$ the mean excitation energy of the atom.
From (\ref{eqn : bethe}) one can see that at low momenta the 
$dE/dx$ falls fast like $1/\beta^2$ (the $1/\beta^2$ region), then 
goes through a minimum and rises very slowly for larger momenta.
For a value of the momentum, $dE/dx$ is distributed according to 
the Landau distribution. 

\begin{figure}[hbtp]
\begin{center}
\mbox{\psfig{figure=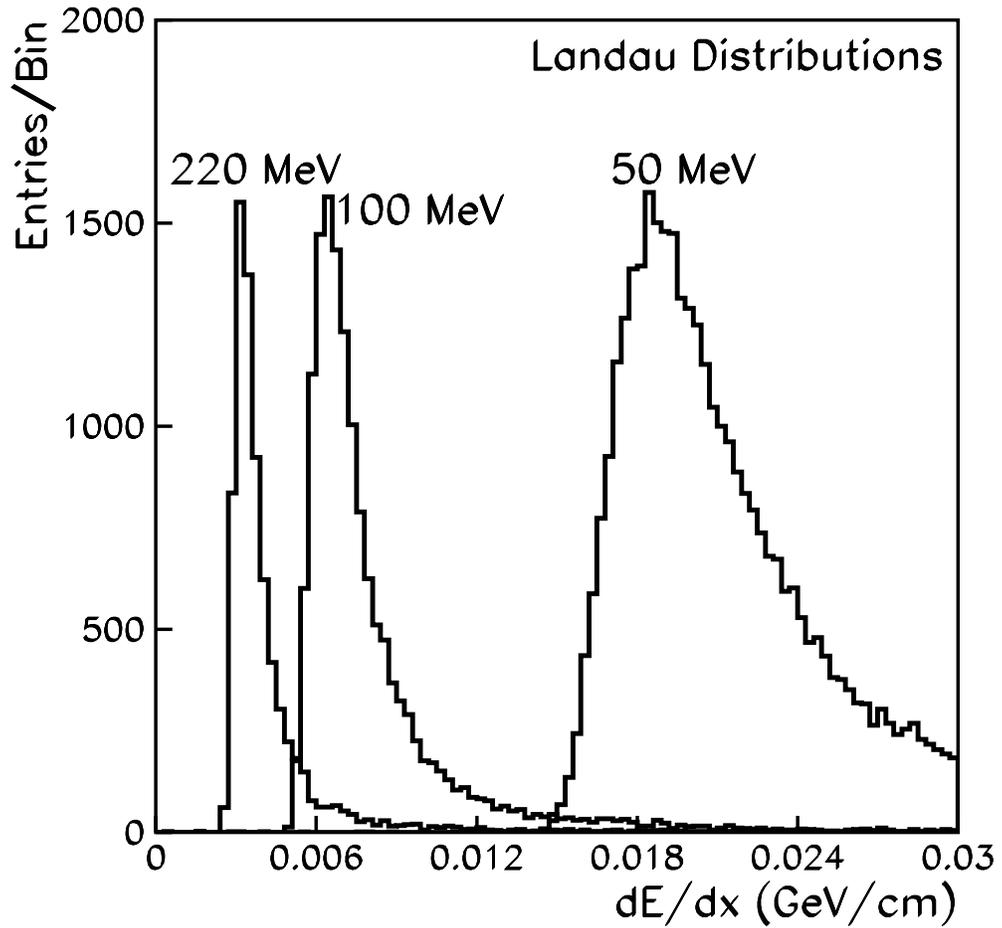,width=5.5in}}
\caption{
Landau $\langle dE/dx \rangle$ 
distributions for pions of momenta 220, 100 and 
50 MeV going through $300\mu m$ of silicon.
}
\label{landau}
\end{center}
\end{figure}

\begin{figure}[hbtp]
\begin{center}
\mbox{\psfig{figure=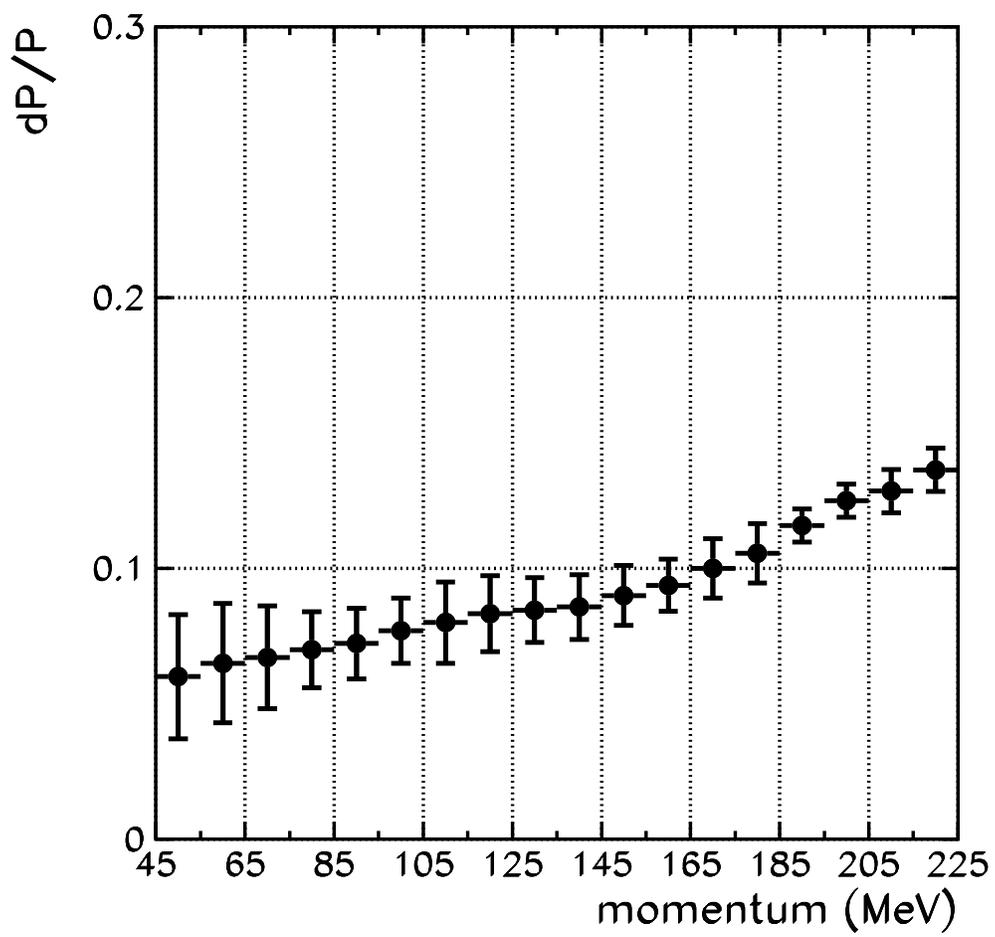,width=5.5in}}
\caption{
Theoretical error in momentum determination.
}
\label{dp_err}
\end{center}
\end{figure}

\noindent
A set of Landau distributions for pions going through $300~\mu m$ 
of silicon with a 
momentum in the range $ 50MeV/c \geq p \geq 220MeV/c$ is 
generated. Some of these distributions are shown in 
fig.~\ref{landau}. 
One can immediately see that the distributions 
are separated in the $1/\beta^2$ region and they get 
wider as the momentum decreases. 
It will be shown that the width of the distributions 
alone induces a relatively small error in the momentum determination 
of the beam. The overlap of the distributions though,  
makes a reliable momentum estimation impossible away from the 
$1/\beta^2$ region because a value of the dE/dx corresponds to 
a momentum range greater than 1 GeV wide. 
Thus it is only in the $1/\beta^2$ region 
where the overlap is minimum that we can apply our method. 
In fig.~\ref{dp_err} we show the momentum error induced by 
the Landau distributions alone (i.e. errors due to overlap 
were not included). 
For a particular momentum the momentum error 
was determined as follows: 
\begin{itemize}
\item
the $dE/dx$ was mapped to a momentum 
using the Bethe-Bloch formula (\ref{eqn : bethe}). A momentum 
distribution was created about the original pion momentum.
\item
We obtained the distribution of the ratio  
$(p_{mapped} - p_{actual})/p_{actual} $
and fitted assuming it follows the Gaussian distribution.
\item
The momentum error plotted in fig.~\ref{dp_err} 
is the standard deviation of the fit:
\begin{eqnarray}
dp/p = \sigma((p_{mapped} - p_{actual})/p_{actual}) = \sigma (\Delta p/p)
\label{eqn : res1}
\end{eqnarray}
We will call (\ref{eqn : res1}) 
the momentum resolution which is an index of the accuracy of 
the particle momentum determination.
\end{itemize}
In fig.~\ref{dp_err} we see that the momentum 
resolution ranges from 6 to 14$\%$ 
in the interesting momentum region. The error bars come from the 
uncertainty that enters our caclulation due to the fit.
As we already mentioned the Landau distributions overlap for 
close enough momenta even in the $1/\beta^2$ region but no 
significant overlap occurs for distributions with momenta 
larger than one standard deviation of (\ref{eqn : res1}). As 
a result the momentum resolution in the $1/\beta^2$ region 
is $\simeq 10\%$.
In this momentum region the momentum resolution obtained from 
standard fitting techniques is dominated by 
the multiple Coulomb scattering and is of order
$20-25\%$ or higher \cite{mar},\cite{bock}, for a few layer 
Silicon based tracking detectors.

The above result comprises the central point
of this paper: for few layer silicon based detectors 
with good hit energy loss $dE$ reconstruction efficiency, the momentum
is more accurately determined from the $dE/dx$ distribution in the 
low $p_\perp$ region ($1/\beta^2$ region of the $dE/dx$ plot) than 
from standard fitting methods.

\section{Simulations and Results}
\vspace{0.2in}

In this section our technique is applied on the STAR-SVT
\cite{cdr}. This
detector consists of three concentric cylindrical layers covered 
with silicon drift detectors of $300 \mu m$ thickness. SVT is capable of 
performing stand-alone tracking using the $x,y,z$ coordinates of the 
3 hits that a charged particle leaves when it crosses the detector.
The tracking efficiency (i.e. the percentage of 
triplets of hits which correspond to the actual tracks that 
crossed the detector) is about $90\%$ for the low momentum region and 
for tracks that originate from the main collision vertex (primary 
tracks). These tracks (in the region below $p_{\perp}=200~MeV/c$)
are mostly pions ($90\%-95\%$). The momentum resolution $dp/p$
of the SVT has been extensively studied 
\cite{mar} and the results show 
that the resolution is rather poor for $p_{\perp}<200~MeV/c$. For 
primary tracks the momentum resolution 
$dp / p \simeq 25\%$ while for secondary 
tracks $dp / p \geq 30\%$. These numbers are in agreement
with estimates for a 3 layer tracking detector from analytical 
empirical formulae \cite{bock}.

\begin{figure}[hbtp]
\begin{center}
\mbox{\psfig{figure=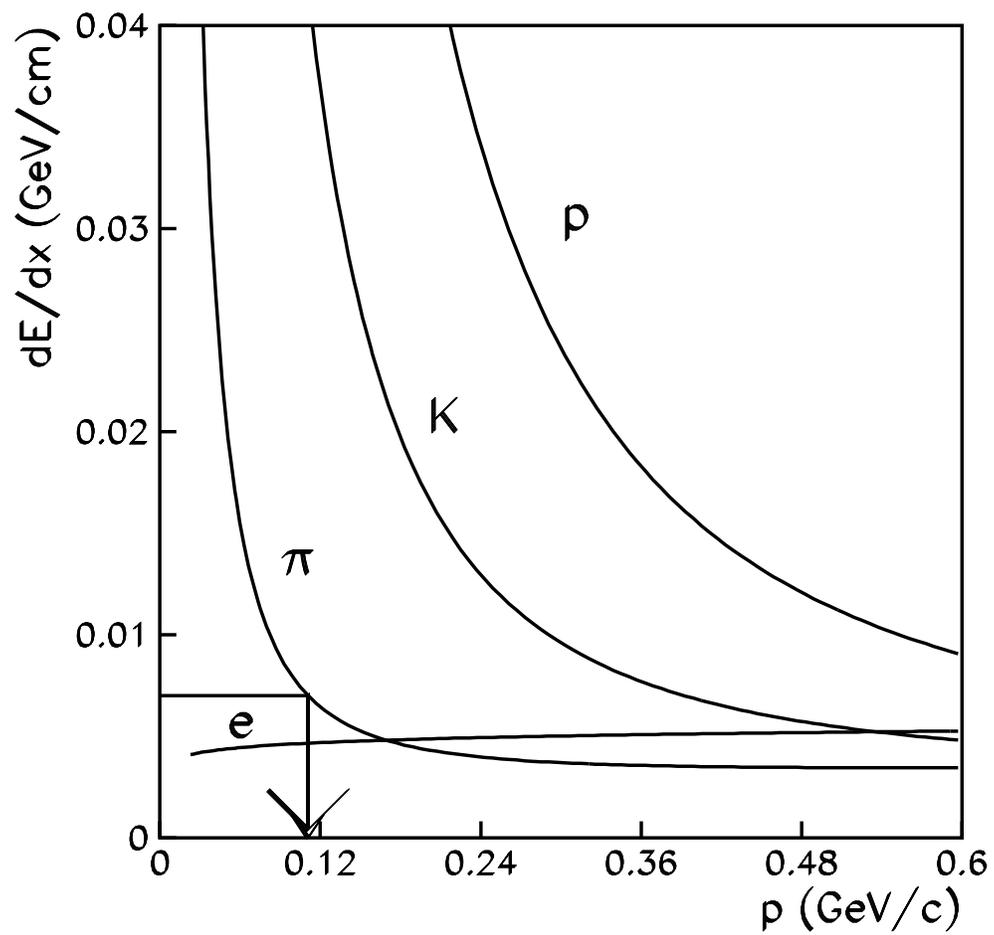,width=5.5in}}
\caption{
$\langle dE/dx \rangle$ as a function of momentum in STAR-SVT.
For low momentum pions one can use the pion curve to extract 
the momentum from the measured $dE/dx$ as shown.
}
\label{dedx_p}
\end{center}
\end{figure}

\noindent
To apply our method in SVT we use the $dE/dx$ information as 
follows:
\begin{enumerate}
\item
We generate the pion $dE/dx$~versus~$p$ band 
using simulations that 
are verified by experimental data. This is done by sending pions 
of known momentum through the detector and extracting their $dE$.
\item
We obtain the Landau distributions $dE/dx(p)$ for various momenta 
using a $50~MeV/c$ momentum step. Every point in these Landau 
distributions corresponds to the truncated average $dE/dx$ of a track.
By truncated we mean that the maximum dE of the three hits is rejected
and that $dE/dx=\frac{dE/dx(1)+dE/dx(2)}{2}$. In this way the long 
tail of the Landau distribution is reduced generating a more 
Gaussian-like distribution.
\item
The mean $dE/dx$ for each Landau distribution is obtained by
performing Gaussian fit (fig.~\ref{dedx_p}).
\item
After we collect a set of mean values for a range of momenta 
from $50~MeV/c$ to $1~GeV/c$, we fit them using a polynomial fit to
obtain an analytical expression for the mean 
\begin{equation}
\langle dE/dx \rangle(p) = Polyn(p).
\label{eqn : mde}
\end{equation}
This expression maps the measured $\langle dE/dx\rangle$ of a track to a 
momentum (fig.~\ref{dedx_p}).
After tracking and standard fitting (typically helix fit) have been 
performed, we replace the momentum of the tracks with $p\leq200~MeV/c$
with the momentum obtained from (\ref{eqn : mde}).
\end{enumerate}

\begin{figure}[hbtp]
\begin{center}
\mbox{\psfig{figure=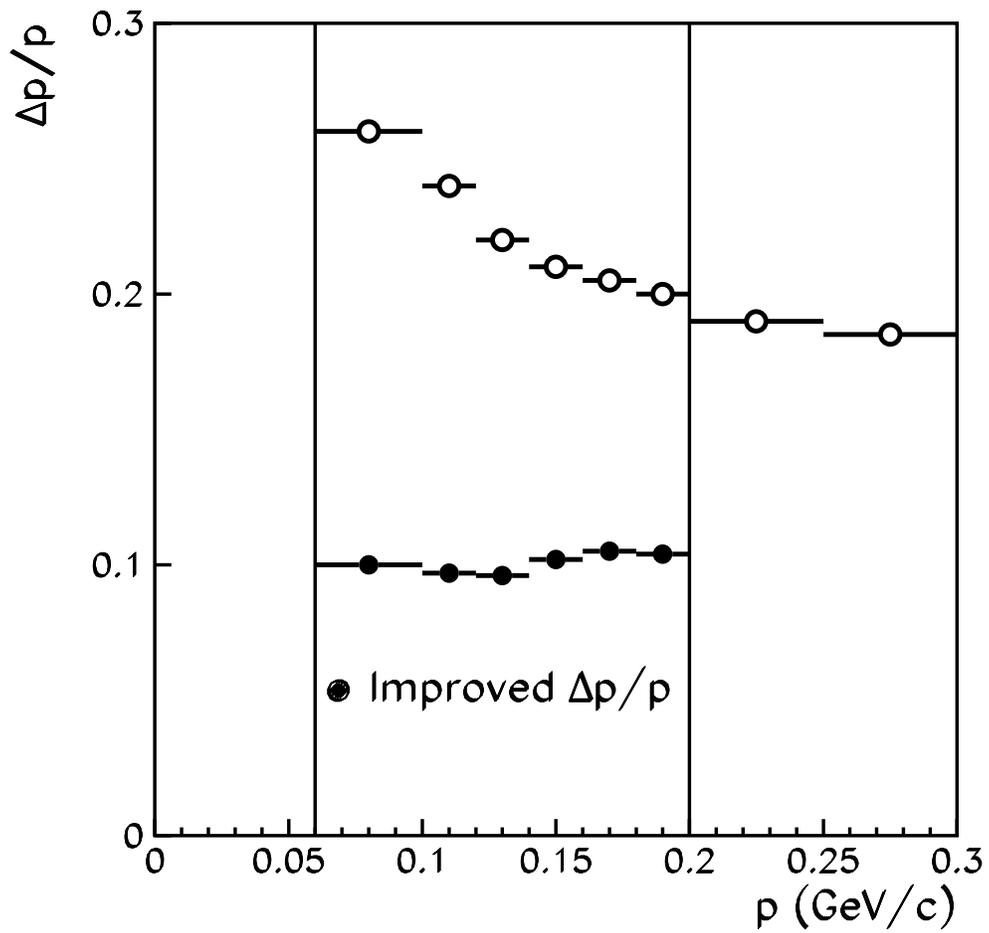,width=5.5in}}
\caption{
Momentum resolution improvement in the momentum 
range $50-200 MeV/c$ in STAR-SVT.
Open circles denote the resolution obtained by 
a helix fit and filled circles the resolution when the momentum 
is extracted from the dE/dx information.
}
\label{svt_pimp}
\end{center}
\end{figure}

Fig.~\ref{svt_pimp} shows the obtained momentum resolution for 
pions. In the momentum region below $200~MeV/c$ one can see that 
the average momentum resolution is $dp/p \simeq 10-11\%$ while
for momenta 
greater than $200~MeV/c$ the resolution is greater than $20\%$ 
because the method cannot be applied there and the momentum 
was obtained from a helix fit.

We expect that our method will be really useful for 
this low momentum region that only the Silicon tracker can be used 
for tracking. Fig.~\ref{svt_pimp} clearly demonstrates that in the 
low momentum region the method provides at least a factor of two 
improvement of the momentum resolution.
In these simulations a perfect knowledge of the energy loss $dE$
in the silicon was assumed. In reality the $dE$ of each hit on a
wafer carries a measurement error which finally affects the 
momentum correction. On the other hand silicon detectors are 
capable of a quite accurate measurement of the $dE$. 
Another problem is the tracking efficiency for low momentum tracks;
when a hit does not belong to a track the corresponding $dE$ 
measurement is wrong and this will affect the momentum correction.
Our simulations include the tracking efficiency errors.

\section{Conclusions}
\vspace{0.2in}
In this paper a new method for improving the momentum determination 
of low momentum particles in silicon trackers was presented. The method
was applied to a heavy ion experiment detector and an 
improvement of the momentum resolution 
by more than a factor of two was found. The method can 
be applied to any silicon detector that provides good $dE$ 
information as in STAR-SVT at RHIC and ALICE at CERN. 
Knowledge of the pion momentum ($p_\perp<200MeV$) 
with an accuracy better than $15\%$ 
can allow for interaction source size measurement using pion correlations, 
collision temperature studies, 
secondary soft pion reconstruction from $V_0$ decays and
searches for pion excesses predicted in the QGP phase at low 
$p_\perp$.


\pagebreak

\end{document}